\documentclass[prx,showpacs,priprent,twocolumn,superscriptaddress]{revtex4-1}
\usepackage{}
\usepackage{mathrsfs}
\usepackage{amsfonts}
\usepackage{txfonts}
\usepackage{amssymb}
\usepackage{graphicx}
\usepackage{bm}
\usepackage{color}
\usepackage[normalem]{ulem}

\newcommand{\ket}[1]{|#1\rangle}
\newcommand{\bra}[1]{\langle #1|}

\begin{document}
\title{Realizing nonadiabatic holonomic quantum computation beyond the three-level setting}
\author{G. F. Xu}
\affiliation{Department of Physics, Shandong University, Jinan 250100, China}
\author{P. Z. Zhao}
\affiliation{Department of Physics, Shandong University, Jinan 250100, China}
\author{Erik Sj\"oqvist}
\email{erik.sjoqvist@physics.uu.se}
\affiliation{Department of Physics and Astronomy, Uppsala University,
Box 516, Se-751 20 Uppsala, Sweden}
\author{D. M. Tong}
\email{tdm@sdu.edu.cn}
\affiliation{Department of Physics, Shandong University, Jinan 250100, China}
\date{\today}
\begin{abstract}
Nonadiabatic holonomic quantum computation (NHQC) provides a method to implement
error resilient gates and that has attracted considerable attention recently. Since it was
proposed, three-level $\Lambda$ systems have become the typical building block for
NHQC and a number of NHQC schemes have been developed based on such systems.
In this paper, we investigate the realization of NHQC beyond the standard three-level setting.
The central idea of our proposal is to improve NHQC by enlarging the Hilbert space of the building block system and letting it have a bipartite graph structure in order to ensure
purely holonomic evolution. Our proposal not only improves conventional qubit-based NHQC
by efficiently reducing its duration, but also provides implementations of qudit-based NHQC.
Therefore, our proposal provides a further development of NHQC that can contribute
significantly to the physical realization of efficient quantum information processors.
\end{abstract}
\maketitle
\date{\today}

\section{Introduction}
Quantum computation is realized by using quantum principles and therefore it processes
information differently from classical computation \cite{Nielsen01}. By utilizing quantum
parallelism, quantum computation enables efficient solutions of certain computational
tasks, like factoring large integers \cite{Shor} and searching unsorted databases
\cite{Grover}. However, practical quantum computation is still challenging and requires
further theoretical and technological development. It is known that the main obstacle to
practical quantum computation is errors caused by either inaccurate manipulation of
quantum systems or interaction with their environment. To address this problem, various
error-resilient models have been proposed. Among these error-resilient models, holonomic
quantum computation plays an important role.

Nonadiabatic holonomic quantum computation (NHQC) \cite{Sjoqvist,Xu} is realized
by using nonadiabatic non-Abelian geometric phases \cite{anandan88}. Since these
phases depend only on the global nature of evolution paths, but not on evolution details,
NHQC provides a geometry-based approach for implementing error-resilient quantum
gates. The most attractive feature of NHQC is that it can be performed at high speed
and simultaneously preserves the geometric robustness against errors. Moreover, its
robustness can be widen by combining it with other error reducing methods
\cite{Xu,Zhang1,Liang,Xue,Sun,Liu2019,Zhu2019}. Due to these
features, NHQC has received considerable attention and various nonadiabatic holonomic
schemes have been put forward
\cite{Sjoqvist,Xu,Abdumalikov,Feng,Arroyo,Zu,Liang,Zhang1,Mousolou,Xue,Xu3,Sjoqvist2,
Herterich,Zhang2,Wang,Sun,Xue2017,Li2017,
Sekiguchi,Zhou2017,Hong2018,Zhao2019,danilin18,Xu2018,Zhang2019,Ramberg,Liu2019,
Zhu2019,Yan2019,Li2020}. In particular, NHQC has been demonstrated experimentally in circuit QED
\cite{Abdumalikov,danilin18,Xu2018,Zhang2019,Yan2019}, NMR \cite{Feng,Li2017,Zhu2019}, NV centers in diamond \cite{Arroyo,Zu,Sekiguchi,Zhou2017}, and trapped ions \cite{Li2020}.

Three-level $\Lambda$ systems have become the typical building block for NHQC since
it was proposed. While impressive progress has been achieved by using this setting,
a natural and important topic is whether one can benefit from realizing NHQC with other
building block systems. The investigation of this topic not only enriches the theory of NHQC,
but may also provide a different way to refine NHQC. Furthermore, with the development
of quantum technologies, the ability to control multi-level quantum systems has been
improved significantly. Recently, diverse experimental platforms including photons, NV
centers, trapped ions, and superconducting circuits have begun to explore multi-level-based
quantum information processing. In particular, the challenges of controlling multi-level systems
have been shown to set no fundamental limitations for high-fidelity multi-level-based quantum
gates \cite{Lanyon,Neeley,Fedorov,Peterer,Svetitsky,Kues,Godfrin,Kiktenko2019,Low2020,
Bianchetti2010,Dolde2014,Senko2015,Naik2017,Imany2019,Luo2019,Wang2020}.
This technological progress also greatly encourages us to investigate the topic of realizing
NHQC with other building block systems.

In this paper, we investigate the realization of NHQC beyond the standard three-level setting. Specifically, we improve NHQC
by enlarging the Hilbert space of the building block system
and letting it have a bipartite graph structure. In this way, not only conventional qubit-based NHQC
can be improved, but also the possibility to realize universal qudit-based NHQC is provided.
We base our proposal on trapped ions, for which all relevant technologies, such as initialisation,
read-out, and controllable interactions, are achievable for designing multi-level-based quantum
computation. Our proposal indicates that enlarging the Hilbert space of the building block system and letting it have an appropriate energy level structure can be a promising direction to develop further NHQC.
In this way, our proposal contributes to the physical realization of efficient and robust quantum
information processors.

\section{The proposal} 
We now demonstrate the first merit of enlarging the Hilbert space of the building block system: the duration of NHQC can be efficiently
reduced. Circuit-based quantum computation, including NHQC, uses the three basic
components, one-, two- and multi-qubit gates, to process information. As the number
of operation steps to realize these components decreases, the duration of the computation
can be reduced. A shorter duration corresponds to a reduced exposure to errors caused by
decoherence and therefore an increased robustness and precision of the computation.
Note that quantum information is entering noisy intermediate-scale quantum era, in
which quantum computers lack the resources for full fault tolerance and therefore can
only support computation of short duration. This makes the reduction of computation
duration a pertinent issue.

\begin{figure}[htbp]
\begin{center}
\includegraphics[width=3.6cm, height=3.0cm]{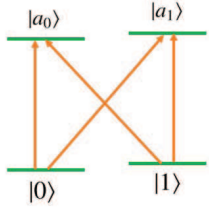}
 \end{center}
\caption{The states in $V_1=\{\ket{0}, \ket{1}\}$ span the computational subspace and
those in $V_2=\{\ket{a_0}, \ket{a_1}\}$ are used as auxiliary states.}
\label{fig1}
\end{figure}

To reduce the duration of NHQC, one may use the building block system having the structure in Fig.~\ref{fig1}. This system
exhibits a bipartite graph structure, i.e., its energy levels are partitioned into two sets
$V_1=\{\ket{0}, \ket{1}\}$ and $V_2 = \{\ket{a_0}, \ket{a_1}\}$, and no transitions
exist within each set. The states in $V_1$ span the computational subspace
and those in $V_2$ are auxiliary states. The key point with the bipartite structure is
that it allows us to perform nonadiabatic gates that are purely holonomic  since there
are no transitions within the computational subspace $V_1$.

We begin by showing the realization of holonomic one-qubit gates with the system having the structure in Fig.~\ref{fig1}. One way for the realization is as follows. Consider one ion and apply resonant laser fields to drive the transitions between one of the
auxiliary states, say $\ket{a_0}$, and the computational states $\ket{0}$ and $\ket{1}$.
In this way, one designs the Hamiltonian
\begin{eqnarray}
H_1(t) = \Omega_1^0(t)\ket{a_0}\bra{0} +
\Omega_1^1(t)\ket{a_0}\bra{1} + {\rm H.c.},
\end{eqnarray}
where $\Omega_1^0(t)$ and $\Omega_1^1(t)$
are Rabi frequencies chosen such that  $\Omega_1^0(t) / \Omega_1^1(t)$ is time independent.
This makes $H_1(t)$ to commute with itself at different times, which is an important ingredient
for guaranteeing the holonomic feature, as it ensures that the dynamical phases
$\int_0^t \bra{k} U^{\dagger} (t',0) H_1 (t') U(t',0) \ket{l} dt'$, $k,l=0,1$ and $U(t',0)$ the
time evolution operator, all vanish (parallel transport). The Hamiltonian $H_1(t)$ describes
an effective three-level $\Lambda$ system, which is known to be sufficient to realize arbitrary
holonomic one-qubit gates \cite{Sjoqvist}. Thus, the system in Fig.~\ref{fig1} is universal on
the single qubit level.

We now show the benefits of using the system to realize holonomic
two-qubit gates. Consider two ions each of which having the level structure shown in
Fig.~\ref{fig1}. We couple these two ions by bichromatic lasers that drive transitions
between the computational and auxiliary states. Under large detuning condition, the
single ion transitions are strongly suppressed, while only the double ion transitions
are allowed due to exchange of vibrational energy between the ions. In this way, the
coupling between the computational state $\ket{ij}$ and the auxiliary state $\ket{a_ma_n}$
can be designed, where $i,j,m,n\in\{0,1\}$. See the methods section for details. Based
on the above coupling mechanism, one can realize the Hamiltonian
\begin{eqnarray}
H_2 (t) = \Omega_2^0 (t) \ket{a_0a_0} \bra{\phi_0} + \Omega_2^1(t) \ket{a_1a_1} \bra{\phi_1} +
{\rm H.c.},
\end{eqnarray}
where $\Omega_2^0(t)$ and $\Omega_2^1(t)$ are Rabi frequencies, and $\ket{\phi_0}$ and
$\ket{\phi_1}$ are states residing in the computational subspace spanned by $\ket{10}$ and
$\ket{11}$. Note that, contrary to the realization of the holonomic one-qubit gate by means
of $H_1 (t)$ above, the ratio of the Rabi frequencies is allowed to be time dependent,
since $\ket{\phi_0}$ and $\ket{\phi_1}$ couple to orthogonal auxiliary states $\ket{a_0a_0}$
and $\ket{a_1a_1}$. We choose the coupling parameters so that $\ket{\phi_0}$ and
$\ket{\phi_1}$ become mutually orthogonal and divide the evolution into two intervals
$0\leq{t}<\tau$ and $\tau\leq{t}\leq{T}$. We divide the run time $T$ so as to satisfy $\int_{t_0}^{t_1}
|\Omega_2^0(t)|dt = \int_{t_0}^{t_1} |\Omega_2^1(t)|dt = \pi/2$ with $t_0=0$, $t_1=\tau$ or
$t_0=\tau$, $t_1=T$, and set the phase of $\Omega_2^0(t)$ (and $\Omega_2^1(t)$) to be
different constants for $0\leq{t}<\tau$ and $\tau\leq{t}\leq{T}$. One thereby obtains the gate
\begin{eqnarray}
U_2 = \ket{00} \bra{00} + \ket{01} \bra{01} + e^{i\gamma_0} \ket{\phi_0} \bra{\phi_0} +
e^{i\gamma_1}\ket{\phi_1}\bra{\phi_1},
\end{eqnarray}
where $\gamma_0$ and $\gamma_1$ are relative phases that can be controlled by varying
the step-wise phase changes of $\Omega_2^0(t)$ and $\Omega_2^1(t)$ during the evolution.
The above realization can be further generalized with the bichromatic coupling mechanism.
Specifically, one can design the Hamiltonian
\begin{eqnarray}
H_3 (t) & = & \Omega_3^{a}(t) \ket{a_0a_0}\bra{\phi_a} +
\Omega_3^{b}(t) \ket{a_1a_1} \bra{\phi_b}
\nonumber \\
 & & + \Omega_3^{c}(t) \ket{a_0a_1} \bra{\phi_c} + {\rm H.c.},
\end{eqnarray}
where $\Omega_3^{a}(t)$, $\Omega_3^{b}(t)$, and $\Omega_3^{c}(t)$ are Rabi
frequencies, and $\ket{\phi_a}$, $\ket{\phi_b}$, and
$\ket{\phi_c}$ are states residing in the space spanned by $\ket{00}$, $\ket{01}$,
$\ket{10}$, and $\ket{11}$. Similar to realizing $U_2$, we choose coupling parameters
so that $\ket{\phi_a}$, $\ket{\phi_b}$, and $\ket{\phi_c}$ become mutually
orthogonal and divide the evolution into two time intervals. Moreover,
the phases of $\Omega_3^{a}(t)$, $\Omega_3^{b}(t)$, and $\Omega_3^{c}(t)$ are
different constants for different intervals. In this way, one realizes the gate
\begin{eqnarray}
U_3=\ket{\phi_\perp}\bra{\phi_\perp}+e^{i\gamma_a}\ket{\phi_a}\bra{\phi_a}
+e^{i\gamma_b}\ket{\phi_b}\bra{\phi_b}+e^{i\gamma_c}\ket{\phi_c}\bra{\phi_c},
\end{eqnarray}
where $\gamma_a$, $\gamma_b$, and $\gamma_c$ are variable relative phases
and $\ket{\phi_\perp}$ is orthogonal to $\ket{\phi_a}$, $\ket{\phi_b}$, and $\ket{\phi_c}$.
The gate $U_2$ represents arbitrary two-qubit controlled gates, while $U_3$ represents
arbitrary two-qubit gates. The holonomic feature of $U_2$ and $U_3$ follows from the
bipartite structure of the underlying four-level systems, but can also be verified explicitly
by checking the parallel transport and cyclic conditions. In contrast, if one uses
three-level ions to realize $U_2$ or $U_3$, two or more sequentially implemented gates
are needed, which increases the number of operation steps of the realization and therefore
the duration of the whole computation. Thus, the system in Fig.~\ref{fig1} brings enhanced
flexibility to realizing two-qubit nonadiabatic holonomic gates and therefore offers benefits
to the reduction of computation duration.

Multi-qubit gates can be built with gates from the universal set of one- and two-qubit gates.
However, this procedure typically becomes very demanding as the number of such gates
rapidly grows with the size of the computational problem. Thus, finding a way to realize
multi-qubit gates with fewer steps is important and a key factor in reducing computation
duration. Among multi-qubit gates, controlled gates play a particularly prominent role. They
are frequently used in various quantum algorithms and quantum error correction schemes
\cite{Nielsen01,Shor,Grover}. We next demonstrate the benefits of using the level structure
in Fig.~\ref{fig1} to realize holonomic multi-qubit controlled gates.

The essence of our method is to separate the ions into blocks and use the freedom
of the four-level system to permit operations acting on different blocks to be implemented
in parallel, so that the number of operation steps can be reduced. Consider $N$ ions confined
in a linear trap, each of which having the bipartite graph structure shown in Fig.~\ref{fig1}.
We divide the $N$ ions into $m\geq1$ blocks and let each block $k\in\{1,\ldots,m\}$
contain $n_k$ ions. We conisder for clarity a specific realization procedure to demonstrate
our method.

Suppose there are two blocks $k=1,2$, each having six ions $(n_1=n_2=6)$ and each ion
encoding a qubit. We consider the realization of a nonadiabatic holonomic controlled phase flip gate that acts on the ions of these two blocks so that $\ket{1\ldots1}$ is taken into $-\ket{1\ldots1}$, while the remaining computational states are kept unchanged.
We denote the ions in each block as $1,\ldots,6$. For block $k=1$, we first implement
the transitions $\ket{11}_{1,2} \rightarrow \ket{a_0a_0}_{1,2}$ and $\ket{01}_{3,4}
\rightarrow \ket{a_1a_1}_{3,4}$; secondly, we implement the transitions
$\ket{a_01}_{2,4} \rightarrow \ket{1a_0}_{2,4}$ and $\ket{01}_{5,6} \rightarrow
\ket{a_1a_1}_{5,6}$; thirdly, we implement the transition $\ket{a_01}_{4,6} \rightarrow
\ket{1a_0}_{4,6}$. For block $k=2$, we first implement the transitions $\ket{11}_{1,2}
\rightarrow \ket{a_0a_0}_{1,2}$, $\ket{01}_{3,4} \rightarrow \ket{a_1a_1}_{3,4}$ and
$\ket{0}_6 \rightarrow\ket{a_1}_6$; the second and third steps are respectively the
same as the second and third steps of block $k=1$. It is noteworthy that the operations
acting on different blocks can be implemented in parallel. Thus, only three steps are
needed for the above operations. The fourth step is to couple the two blocks by implementing
$\ket{a_0a_0}_{6_1,6_2} \rightarrow -\ket{a_0a_0}_{6_1,6_2}$ with the help of the intermediate
state $\ket{00}_{6_1,6_2}$; here, we have used the notation $\mu_k$ to denote ion $\mu$
of block $k$ for clarity. Finally, we implement the first three steps again but with both the
implementation order and the implemented transitions reversed. Thus, the realization costs
$7$ steps and the realized gate can be verified to yield a holonomic controlled phase gate.
See the methods section for further details.

The above procedure can be generalized to the case of more than two blocks. For
illustrative purpose, consider three blocks $k=1,2,3$, each block containing six ions. The
first three steps of blocks $k=1,2$ are the same as before, while the first three steps of
block $k=3$ are the same as those of block $k=2$. Thus, because of parallelism, it costs
three steps in total. The fourth step is to couple the first two blocks by implementing
$\ket{a_0a_0}_{6_1,6_2} \rightarrow \ket{00}_{6_1,6_2}$. Next, we couple the second
and third blocks by implementing $\ket{0a_0}_{6_2,6_3} \rightarrow -\ket{0a_0}_{6_2,6_3}$
with the help of state $\ket{a_00}_{6_2,6_3}$. Finally, we repeat the first
four steps but with the implementation order and the implemented transitions reversed.
The above realization costs $9$ steps and the realized
controlled phase gate is purely holonomic. Generalizing to four such blocks requires
$11$ steps. In fact, the parallelism offered by four-level systems permits the
blocks to contain more ions: in the three-block case, the third block can contain eight
ions; in the four-block case, the third and fourth blocks can respectively contain eight
and ten ions. Although the number of ions increases, the number of operation steps
remain $9$ and $11$, respectively.

The four-level system allows us to use $7$, $9$, and $11$ steps to realize holonomic
controlled phase gates acting on various number of qubits. In contrast, to achieve
the same task by using three-level $\Lambda$ systems, the best known scheme needs
$21$, $37$, and $57$ steps, respectively \cite{Zhao2019}. Thus, the number of steps
can be reduced significantly by replacing the standard $\Lambda$ setting by the four-level
building block. From the above comparison, one can also see that the reduction of the
number of steps becomes more obvious with the increase of the number of blocks, i.e.,
the number of qubits the controlled phase gate acts on. Note that for a practical quantum
computer, the number of qubits is usually very large. In this case, the number of steps
may also be saved by orders of magnitude. It is also noteworthy that although we use
a specific realization procedure, our method can be generalized to realize holonomic
controlled gates acting on different number of qubits.

We have shown how NHQC can be improved by using a building block system with a four-level bipartite graph structure. We next take this idea one step further and show that by enlarging the Hilbert space of the building block system and letting it have a bipartite graph structure, one can have a feasible platform to develop qudit-based NHQC, in which the holonomic gates act on $d$-dimensional quantum objects. We explicitly demonstrate the realization of qutrit-based ($d=3$) NHQC to show this.

Before proceeding further, we briefly explain the necessity of developing qudit-based NHQC. Compared to qubit settings, qudit-based processors can store exponentially larger amount of information and thereby provide reduction of the circuit complexity and simplification of the experimental setup \cite{Review1,Review2,Review3}. Qudit-based processors can also enable enhancement of the algorithm efficiency, favorable error thresholds and high-fidelity magic-state distillation \cite{Gokhale,Campbell,Campbell1,Andrist,Michael2016,Muralidharan2017}.
This benefits quantum error corrections and is essential for scalable quantum computation.
Particularly, qudit-based processors have recently begun to be explored on various experimental
platforms \cite{Bianchetti2010,Dolde2014,Senko2015,Naik2017,Imany2019,Luo2019,Wang2020}.
Because of these features, developing qudit-based NHQC is of significance.

\begin{figure}[htbp]
\begin{center}
\includegraphics[width=4.2cm, height=3.0cm]{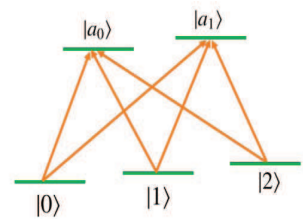}
 \end{center}
\caption{The states in $V_1=\{\ket{0}, \ket{1}, \ket{2}\}$ span the computational subspace
and those in $V_1=\{\ket{a_0}, \ket{a_1}\}$ are used as auxiliary states.}
\label{fig2}
\end{figure}

We now explicitly develop qutrit-based ($d=3$) NHQC with ions comprising the five-level bipartite
graph structure shown in Fig.~\ref{fig2}. The states $\ket{0}$, $\ket{1}$, $\ket{2}$ span the qutrit
computational subspace and $\ket{a_0}$, $\ket{a_1}$ are auxiliary states. We show that the
system in Fig.~\ref{fig2} permits us to conveniently realize not only holonomic one-qutrit and
two-qutrit gates, but also holonomic multi-qutrit controlled gates.

To realize holonomic one-qutrit gates, we apply resonant laser fields to drive transitions
between the computational and auxiliary states, as described by the Hamiltonian
$H(t) = \sum_{i=0}^2 \Omega_{0i} (t) \ket{a_0} \bra{i} + \sum_{j=0}^2 \Omega_{1j} (t)
\ket{a_1}\bra{j} + {\rm H.c.}$, where $\Omega_{0i}(t)$ and $\Omega_{1j}(t)$ are Rabi
frequencies. By letting $\Omega_{0i}(t)/\Omega_{0j}(t)$ and $\Omega_{1i}(t)/\Omega_{1j}(t)$
be constant for all pairs $i,j=0,1,2$, the Hamiltonian turns into
\begin{eqnarray}
H_4 (t) = \Omega_4^0 (t) \ket{a_0} \bra{\psi_0} + \Omega_4^1 (t) \ket{a_1} \bra{\psi_1} +
{\rm H.c.},
\end{eqnarray}
where $\Omega_4^0(t)$ and $\Omega_4^1(t)$ are common envelopes, and $\ket{\psi_0}$
and $\ket{\psi_1}$ are states residing in $L=\text{Span}\{\ket{0},\ket{1},\ket{2}\}$. By choosing
the pulse areas appropriately, one can use $H_4(t)$ to generate
\begin{eqnarray}
U_4=\ket{\psi_\perp}\bra{\psi_\perp}+e^{i\xi_0}\ket{\psi_0}\bra{\psi_0}
+e^{i\xi_1}\ket{\psi_1}\bra{\psi_1},
\end{eqnarray}
where $\ket{\psi_0}$, $\ket{\psi_1}$, and $\ket{\psi_\perp}$ can be chosen mutually
orthogonal, and $\xi_0$ and $\xi_1$ are relative phases. The gate $U_4$ represents
an arbitrary one-qutrit gate acting on $L$ and can be verified to be holonomic. Since
$U_4$ is a gate acting on a three-dimensional Hilbert space, one needs to use two
one-qubit gates and an entangling two-qubit gate to simulate it with qubit-based NHQC.

To realize holonomic two-qutrit gates, we consider two ions each of which has the structure
in Fig.~\ref{fig2}. We use the bichromatic laser pulse mechanism to couple the two ions,
just as in the realization of $H_2(t)$ and $H_3(t)$ above, and consider
\begin{eqnarray}
H_5(t) & = & \Omega_5^{a}(t) \ket{a_0a_0} \bra{\psi_a} +
\Omega_5^{b}(t) \ket{a_1a_1} \bra{\psi_b}
\nonumber \\
 & & + \Omega_5^{c}(t) \ket{a_0a_1} \bra{\psi_c} + {\rm H.c.},
\end{eqnarray}
where $\Omega_5^{a}(t)$, $\Omega_5^{b}(t)$, and $\Omega_5^{c}(t)$ are Rabi
frequencies, and $\ket{\psi_a}$, $\ket{\psi_b}$, and $\ket{\psi_c}$
are mutually orthogonal states residing in the computational subspace
$\text{Span}\{\ket{20}, \ket{21}, \ket{22}\}$. By following the approach that results in $U_2$
and $U_3$ above, $H_5(t)$ can be used to realize the two-qutrit gate
\begin{eqnarray}
U_5=I_{\perp}+e^{i\xi_a}\ket{\psi_a}\bra{\psi_a}
+e^{i\xi_b}\ket{\psi_b}\bra{\psi_b}+e^{i\xi_c}\ket{\psi_c}\bra{\psi_c},
\end{eqnarray}
where $I_\perp$ is the identity operator acting on the space orthogonal to $\text{Span}
\{\ket{20}, \ket{21}, \ket{22}\}$, and $\xi_a$, $\xi_b$, and $\xi_c$ are variable
relative phases. $U_5$ is a two-qutrit controlled gates and its holonomic feature
can be verified. It acts on a nine-dimensional space so that one needs to use four qubits
to simulate it with qubit-based NHQC.

We now demonstrate the realization of holonomic multi-qutrit controlled gates with systems
in Fig.~\ref{fig2}. In fact, the method we used to realize holonomic multi-qubit controlled
gates as described above, can be translated to the multi-qutrit case. We use a specific
realization procedure to demonstrate the basic principle. We consider realizing
controlled phase gate acting on qutrits. Here, the controlled phase gate flips the
sign of $\ket{2\ldots2}$, while all other computational states remain unchanged.
To realize this gate, we need to substitute the state $\ket{1}$ in
the previous procedure with the state $\ket{2}$. For example, the transitions
$\ket{11}_{1,2} \rightarrow \ket{a_0a_0}_{1,2}$ and $\ket{a_01}_{2,4} \rightarrow
\ket{1a_0}_{2,4}$ should be replaced by the transitions  $\ket{22}_{1,2} \rightarrow
\ket{a_0a_0}_{1,2}$ and $\ket{a_02}_{2,4} \rightarrow \ket{2a_0}_{2,4}$, respectively.
Furthermore, we need to replace the transition $\ket{01}_{\alpha,\beta}\rightarrow
\ket{a_1a_1}_{\alpha,\beta}$ in the previous procedure with the two transitions
$\ket{02}_{\alpha,\beta} \rightarrow \ket{a_1a_1}_{\alpha,\beta}$ and $\ket{12}_{\alpha,\beta}
\rightarrow \ket{a_0a_1}_{\alpha,\beta}$. With these changes, one can realize the desired
holonomic controlled phase gates acting on qutrits. The above procedure can also be
generalized to realize arbitrary holonomic multi-qutrit controlled gates. Thus, the graph
structure shown in Fig.~\ref{fig2} permits an efficient way to realize holonomic multi-qutrit
controlled gates.

We have shown that various holonomic qutrit-based gates can be conveniently realized
with the system in Fig.~\ref{fig2}, which benefits the reduction of the duration of NHQC.
Moreover, since qutrits store exponentially larger amount of information than qubits,
qutrit-based NHQC has the advantages mentioned above, such as reduction of the circuit
complexity and simplification of the experimental setup, etc.

The idea of our proposal is to improve NHQC by enlarging the Hilbert space of the building block system and letting it have a bipartite graph structure, and we have demonstrated the cases of
four-level and five-level bipartite graph systems described in Figs.~\ref{fig1} and \ref{fig2}
respectively. We can take the idea one more step further to consider a bipartite graph
system with more energy levels as the building block system for NHQC. In this case,
qudit-based NHQC with larger computational space can be developed and the specific
realization procedure is similar to that for realizing holonomic qutrit-based gates
demonstrated above.

\section{Conclusion}
In this work, we propose settings for NHQC beyond the typical three-level
$\Lambda$ configuration. The proposed settings have a larger Hilbert space and meanwhile
a bipartite graph structure. Our results show that using the proposed settings not only
improves conventional qubit-based NHQC by efficiently reducing its duration, but also
provides a platform for efficiently realizing universal qudit-based NHQC. Our proposal
opens up for several extensions. First, while our proposal uses bipartite graph systems,
other kinds of multi-level systems may also be useful for refining NHQC. The investigation
of such schemes may provide a new framework to develop NHQC. Secondly, while our
proposal provides one way of realizing holonomic gates with bipartite graph multi-level
systems, investigating other coupling sequences for realizing holonomic gates with such
systems is worth paying attention to. Such an investigation may result in improved schemes
to use bipartite graph systems for NHQC. Finally, qudit-based NHQC can be combined
with other error-resilient methods, such as decoherence-free subspaces \cite{lidar98} and noiseless subsystems \cite{knill00}, to realize quantum information processing with improved robustness features.

\section*{Acknowledgments}
This work was supported by the National Natural Science Foundation of China through Grant No. 11775129. P.Z.Z. acknowledges support from the National Natural Science Foundation of China through Grant No. 11947221. E.S. acknowledges support from
the Swedish Research Council (VR) through Grant No. 2017-03832.

\subsection*{Appendix A: The coupling mechanisms}
We demonstrate two mechanisms to couple the
internal states of two ions, labeled as $1$ and $2$, for which all the needed Hamiltonians of our proposal can be realized. Our first coupling mechanism makes use of the transition $\ket{k}\leftrightarrow\ket{a}$ of ion $1$, driven by a red detuned laser with detuning $-(\nu+\delta)$ and Rabi frequency $\omega_{1}(t)$ and by a blue detuned laser with detuning $(\nu-\delta)$ and Rabi frequency $\omega_{2}(t)$, where $\ket{k}$ is one of the computational state and $\ket{a}$ is one of the auxiliary states. Meanwhile, the
transition $\ket{l} \leftrightarrow \ket{a}$ of ion $2$ is driven by a blue detuned laser
with detuning $\nu+\delta$ and Rabi frequency $\omega_{3}(t)$, and $\ket{k}\leftrightarrow
\ket{a}$ of ion $2$ by a red detuned laser with detuning $-(\nu-\delta)$ and Rabi frequency
$\omega_{4}(t)$, where $\ket{l}$ is another computational state. In the above, $\nu$ is the
phonon frequency and $\delta$ is an additional detuning. In this way, one realizes the Hamiltonian
$H(t) = i\eta \omega_1 (t) e^{-i\delta{t}}b^{\dag}\ket{a}_{11}\bra{k} + i\eta\omega_2 (t)
e^{-i\delta{t}} b \ket{a}_{11} \bra{k} + i\eta\omega_3 (t) e^{i\delta{t}} b \ket{a}_{22}
\bra{l} + i\eta\omega_4(t) e^{i\delta{t}} b^{\dag}\ket{a}_{22}\bra{k}$+H.c., where $b$ and
$b^\dag$ are the annihilation and creation operators of the vibrational mode, and $\eta$ is the
Lamb-Dicke parameter that satisfies $\eta^2(n_\nu+1)\ll1$ with $n_\nu$ being the
quantum number of the vibrational mode. If the large detuning condition
$|\delta | \gg | \eta \omega_j (t)|$ is satisfied, the above Hamiltonian can be reduced to
$H_{\rm eff}(t)=\Omega_{kl}(t) \ket{aa}\bra{kl}+\Omega_{kk}(t) \ket{aa} \bra{kk} +
{\rm H.c.}$, where $\Omega_{kl}(t) = - \eta^{2} \omega_{1}(t) \omega_{3}(t) / \delta$ and
$\Omega_{kk}(t) =\eta^{2} \omega_{2}(t) \omega_{4}(t) / \delta$.

We also make use of a second form of coupling mechanism. Here, a pair of laser beams, one of
which with detuning $-(\nu+\delta)$ and the other with detuning $(\nu-\delta)$, are applied
to drive the transition $\ket{k}\leftrightarrow\ket{a}$ of ion $1$; another pair of laser
beams, one of which with detuning $-(\nu+\delta)$ and the other with detuning $(\nu-\delta)$,
to respectively drive the transitions $\ket{l}\leftrightarrow\ket{a}$ and $\ket{k}\leftrightarrow\ket{a}$.
Under the large detuning condition and with the derivation similar to that in the previous
paragraph, one obtains the effective Hamiltonian $H^\prime_{\rm eff} (t) =
\Omega_{al}(t) \ket{ka} \bra{al} + \Omega_{ak} (t) \ket{ka} \bra{ak} + {\rm H.c.}$.

\subsection*{Appendix B: The relevant Hamiltonians}
We now demonstrate the relevant Hamiltonians
used in realizing holonomic multi-qubit controlled phase flip gate. Taking these Hamiltonians
as a reference, one obtains the relevant Hamiltonians used in other realization procedures,
including the realization of holonomic multi-qudit gate. We consider the case where there are
two blocks $k=1,2$, each having six ions with each ion encoding a qubit. We label the ions
as $1,\cdots,6$ in each block. For block $k=1$, to implement the transitions $\ket{11}_{1,2}
\rightarrow \ket{a_0a_0}_{1,2}$ and $\ket{01}_{3,4}\rightarrow\ket{a_1a_1}_{3,4}$, we
use the Hamiltonian $H(t) = \Omega_{1,2}(t)\ket{a_0a_0}_{1,2}\bra{11} + \Omega_{3,4}(t)
\ket{a_1a_1}_{3,4} \bra{01} + {\rm H.c.}$ and the evolution time $T_1$ satisfies $\int_0^{T_1}|
\Omega_{1,2}(t)| dt = \int_0^{T_1} |\Omega_{3,4}(t)| = \pi/2$. The transitions
$\ket{a_01}_{2,4}\rightarrow\ket{1a_0}_{2,4}$ and $\ket{01}_{5,6}
\rightarrow \ket{a_1a_1}_{5,6}$ are implemented by using the Hamiltonian $H(t) =
\Omega_{2,4}(t) \ket{1a_0}_{2,4}\bra{a_01} + \Omega_{5,6} (t) \ket{a_1a_1}_{5,6} \bra{01} +
{\rm H.c.}$ and evolution time $T_2$ satisfying $\int_0^{T_2}|\Omega_{2,4}(t)|dt = \int_0^{T_2}|
\Omega_{5,6}(t)|=\pi/2$. Finally, we implement the transition $\ket{a_01}_{4,6} \rightarrow
\ket{1a_0}_{4,6}$ by using the Hamiltonian $H(t)=\Omega_{4,6}(t)\ket{1a_0}_{4,6}
\bra{a_01} + {\rm H.c.}$ with evolution time $T_3$ satisfying $\int_0^{T_3}
|\Omega_{4,6}(t)| = \pi/2$. For block $k=2$, to implement the transitions
$\ket{11}_{1,2}\rightarrow\ket{a_0a_0}_{1,2}$, $\ket{01}_{3,4}\rightarrow\ket{a_1a_1}_{3,4}$
and $\ket{0}_6\rightarrow\ket{a_1}_6$, we can use the Hamiltonian
$H(t) = \Omega_{1,2}^\prime(t)\ket{a_0a_0}_{1,2}\bra{11} + \Omega_{3,4}^\prime(t)
\ket{a_1a_1}_{3,4} \bra{01} + \Omega_6^\prime(t) \ket{a_1}_6 \bra{0} + {\rm H.c.}$ with
the evolution time $T_1^\prime$ satisfying $\int_0^{T_1^\prime}| \Omega_{1,2}^\prime(t)| dt =
\int_0^{T_1^\prime} |\Omega_{3,4}^\prime(t)| dt = \int_0^{T_1} |\Omega_{6}^\prime(t)| dt =
\pi/2$. The Hamiltonians used in the second and third steps of block $k=2$ are respectively
the same as those used in the second and third steps of block $k=1$. The above operations
cost three steps. After the above operations, we couple the two blocks by implementing
the transition $\ket{a_0a_0}_{6_1,6_2}\rightarrow{-\ket{a_0a_0}_{6_1,6_2}}$, where state
$\ket{00}_{6_1,6_2}$ is the intermediate state and we have used the notation $\mu_k$
to denote ion $\mu$ of block $k$. To implement this operation, we use the Hamiltonian
$H(t)=\Omega_{6_1,6_2}(t)\ket{00}_{6_1,6_2}\bra{a_0a_0}+{\rm H.c.}$ with the evolution
time satisfying $\int_0^{T_4}|\Omega_{6_1,6_2}(t)|=\pi$. We thereafter repeat the operations
of the first three steps, but with both the implementation order and the implemented transitions
reversed, which completes the realization. The used Hamiltonians are the same as those
used in the first three steps. All the above mentioned Hamiltonians can be realized by using
the two coupling mechanisms introduced in the previous subsection. It is noteworthy that these
Hamiltonians always couple states in the computational space with
auxiliary states during the gate realization procedure. This fact guarantees that the realized gate
is purely holonomic.

\end{document}